 \documentclass[final,5p,times,twocolumn]{elsarticle}

\usepackage{amssymb}

\usepackage{journals}

\journal{High Energy Density Physics}

\begin{document}

\begin{frontmatter}



\title{An Experimental Platform for Creating White Dwarf Photospheres in the Laboratory}

\author[UT,Sandia]{Ross~E. Falcon\fnref{NPSC}} 
\ead{cylver@astro.as.utexas.edu}
\author[Sandia]{G.~A. Rochau}
\author[Sandia]{J.~E. Bailey}
\author[UT]{J.~L. Ellis}
\author[Sandia]{A.~L. Carlson}
\author[UT]{T.~A. Gomez}
\author[UT]{M.~H. Montgomery}
\author[UT]{D.~E. Winget}
\author[UT]{E.~Y. Chen}
\author[Sandia]{M.~R. Gomez}
\author[Sandia]{T.~J. Nash}

\address[UT]{Department of Astronomy and McDonald Observatory, University of Texas, Austin, TX 78712, USA}
\address[Sandia]{Sandia National Laboratories, Albuquerque, NM 87185-1196, USA}
\fntext[NPSC]{National Physical Science Consortium Graduate Fellow}

\begin{abstract}

We present an experimental platform for measuring hydrogen Balmer emission and absorption line profiles for plasmas with white dwarf (WD) photospheric conditions ($T_{\rm e}\sim1$\,eV, $n_{\rm e}\sim10^{17}$\,cm$^{-3}$).  These profiles will be used to benchmark WD atmosphere models, which, used with the spectroscopic method, are responsible for determining fundamental parameters (e.g., effective temperature, mass) for tens of thousands of WDs.  Our experiment, performed at the {\it Z} Pulsed Power Facility at Sandia National Laboratories, uses the large amount of x-rays generated from a z-pinch dynamic hohlraum to drive plasma formation in a gas cell.  The platform is unique compared to past hydrogen line profile experiments in that the plasma is radiation-driven.  This decouples the heating source from the plasma to be studied in the sense that the radiation temperature causing the photoionization is independent of the initial conditions of the gas.  For the first time we measure hydrogen Balmer lines in {\it absorption} at these conditions in the laboratory for the purpose of benchmarking Stark-broadened line shapes.  The platform can be used to study other plasma species and to explore non-LTE, time-dependent collisional-radiative atomic kinetics.

\end{abstract}


\end{frontmatter}


\section{Introduction}\label{intro}

The history of laboratory experiments relevant to white dwarf (WD) photospheres follows the progress of theoretical line shapes, particularly for Stark-broadened hydrogen.  During the 1960s and early 1970s, the frequency of shock-heated \cite[e.g.,][]{Berg62,McLean65,Bengtson69}, pulsed discharge \cite[e.g.,][]{Vujnovic62,Hill67,Morris68}, and stabilized arc experiments \cite[e.g.,][]{Wiese63,Shumaker68,Wiese72} was high, tracking with the dynamic advances in theory 
\cite[e.g.,][]{Kolb58,Baranger58,Griem59,Griem62,Kepple68,Smith69}.  As time went on, improving the precision of diagnostic techniques contributed more to the experimental motivation \cite[e.g.,][]{Baessler80,Helbig81} in addition to testing the latest theory \cite[e.g.,][]{Vidal73}.

Astronomical observation frequently uses line shape theory -- white dwarf photospheres in particular are an ideal astrophysical environment for its application \cite[e.g.,][]{Dimitrijevic11}.  Until now, however, this field has not been a significant motivator for experiments.  With new advances in theory \cite[e.g.,][]{Tremblay09,Santos12}, new experimental capabilities available at modern facilities, and a wealth of spectroscopic observations that did not exist much more than a decade ago \cite[e.g.,][]{Kepler07}, this is changing.

\subsection{Motivation}

In WD astronomy, the method of inferring photospheric conditions from the comparison of line profiles in observed spectra to those in synthetic spectra from WD atmosphere models is known as the spectroscopic method \cite[e.g.,][]{Bergeron92b}.  These atmospheric parameters are, namely, the effective temperature $T_{\rm eff}$ and the surface gravity $g$.  The effective temperature of a star is the blackbody temperature defined according to the Stefan-Boltzmann Law so that
\begin{equation}
L=4\pi R^2 \sigma T_{\rm eff}^4,
\end{equation}
where $L$ is the luminosity of the star, $R$ is the stellar radius, and $\sigma$ is the Stefan-Boltzmann constant.  The surface gravity is the gravitational acceleration at the surface of the star and is defined as 
\begin{equation}
g=\frac{GM}{R^2},
\end{equation}
where $G$ is the gravitational constant, and $M$ is the stellar mass.  The surface gravity is conventionally expressed in logarithmic units (log\,$g$).

The spectroscopic method is the most widely-used technique and is responsible for determining parameters for tens of thousands of WDs \cite[e.g.,][]{Liebert05,Kepler07,Koester09b,Castanheira10}.  Atmospheric parameters form the foundation (and limit the ultimate accuracy) of many other research areas, such as determining the age of the universe \cite{Winget87}, constraining the mass of supernova progenitors \cite[e.g.,][]{Williams09}, and probing properties of dark matter axions \cite{Bischoff-Kim08}.

While this method is powerful and precise ($\frac{\delta T_{\rm eff}}{T_{\rm eff}}\sim5$\,\% and $\frac{\delta{\rm log}g}{{\rm log}g}\sim1$\,\% are typical for a given star), its {\it accuracy} is suspect.  Fitting spectra of WDs with $T_{\rm eff}\lesssim12,000$\,K is well-known to be particularly problematic \cite{Bergeron07,Koester09a}.  Log\,$g$ values determined for these cooler WDs go up with decreasing $T_{\rm eff}$, implying an unphysical increase in mean mass that is not found in photometric \cite{Kepler07,Engelbrecht07} or gravitational redshift \cite{Falcon10} studies.  Hypotheses for these discrepancies include that the WD models used in the spectroscopic method may be flawed or contain uncertain input physics \cite[e.g.,][]{Koester09a}.  For example, inaccurate line profiles could affect the inferred WD atmospheric parameters.  Furthermore, the mean mass of WDs determined from spectroscopic investigations disagrees with the mean mass determined from the atmosphere model-independent technique which uses gravitational redshifts \cite{Falcon10}, indicating an underestimated spectroscopic mass at all $T_{\rm eff}$.

WD atmosphere models are still advancing.  The latest models of \citet{Tremblay09} use newly-calculated Stark-broadened line profiles of hydrogen.  In re-analyzing the hydrogen-atmosphere WDs from the Palomar-Green Survey \cite{Liebert05}, their spectroscopic fits yield significant systematic increases in $T_{\rm eff}\sim200-1000$\,K and in log\,$g\sim0.04-0.1$.  This work demonstrates that modified hydrogen line profiles can indeed greatly impact the interpretation of WD spectroscopic observations.  The experimental basis for the line profile theories, however, does not presently provide accurate enough constraints to discern which theoretical model is optimal, if any, and at which conditions.

The differences in the new theory and the resulting impact on the WD atmospheric parameters motivate us to perform new experiments to measure hydrogen Balmer line shapes \cite{Falcon10b}.  In this paper we describe the experimental platform for measuring these lines for H plasmas with WD photospheric conditions 
($T_{\rm e}\sim1$\,eV, $n_{\rm e}\sim10^{17}$\,cm$^{-3}$) and present example data.  We also discuss how these experiments differ from past investigations and the advantages that offers; this includes using the unique x-ray capability of the {\it Z} Pulsed Power Facility \cite{Matzen05} at Sandia National Laboratories.  In a future paper, we will describe detailed results and address their applications in astrophysics, particularly in WD astronomy.

\section{Experimental Setup}

\begin{figure}[]
\includegraphics[width=\columnwidth]{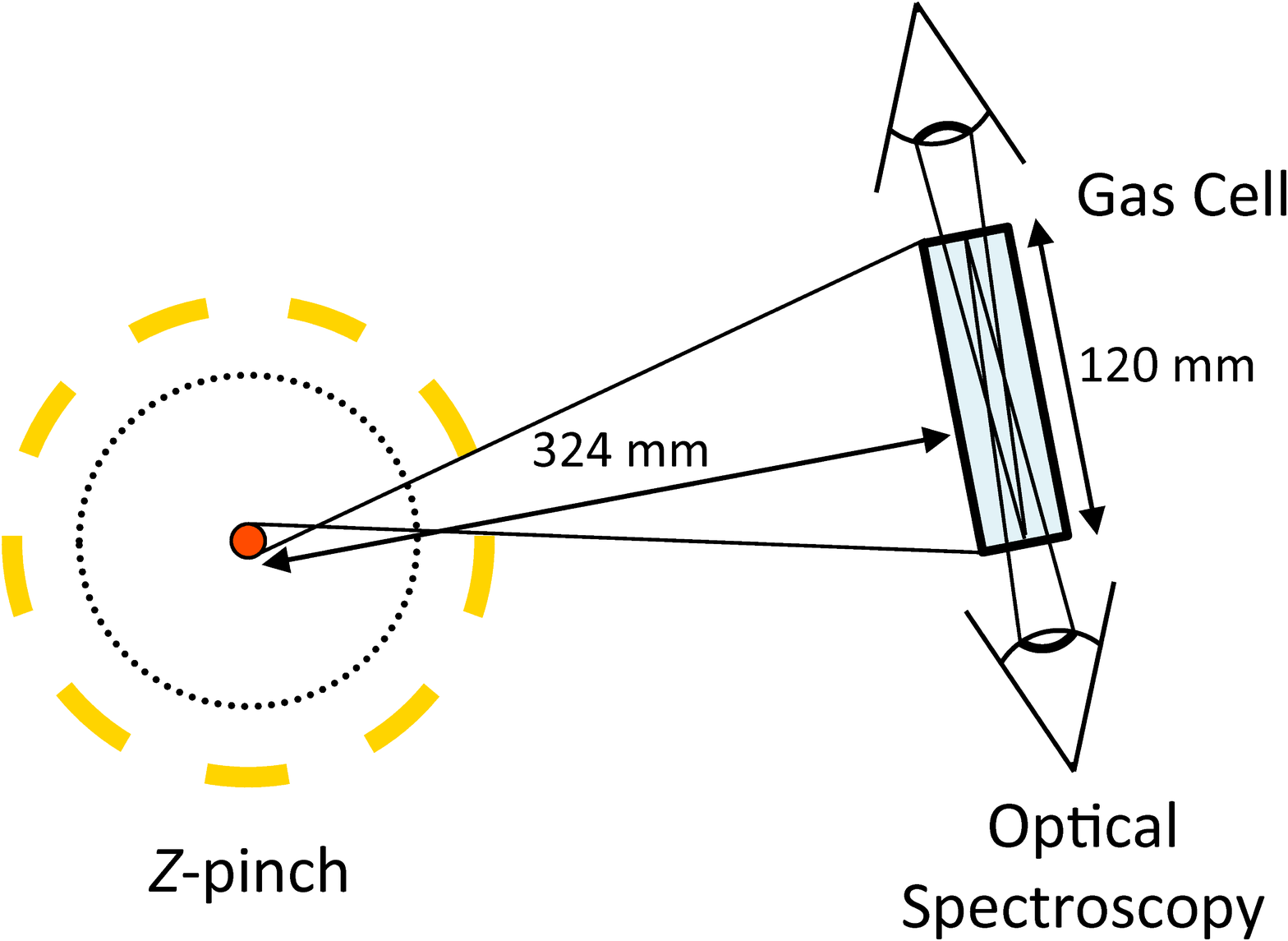}
\caption{Top view schematic of the experimental setup inside the vacuum chamber.  The gas cell sits 324\,mm away from the {\it z} pinch.  We observe the hydrogen plasma along lines of sight perpendicular to the x-rays.\label{overview}}
\end{figure}

Our experiment is part of the {\it Z} Astrophysical Plasma Properties (ZAPP) Collaboration \cite{Montgomery12}.  ZAPP experiments are conducted at the {\it Z} Pulsed Power Facility \cite{Matzen05} at Sandia National Laboratories, and they are performed simultaneously in a large (2200\,ft$^3$) vacuum chamber making use of the same {\it z}-pinch x-ray source.

\begin{figure}[]
\includegraphics[width=\columnwidth]{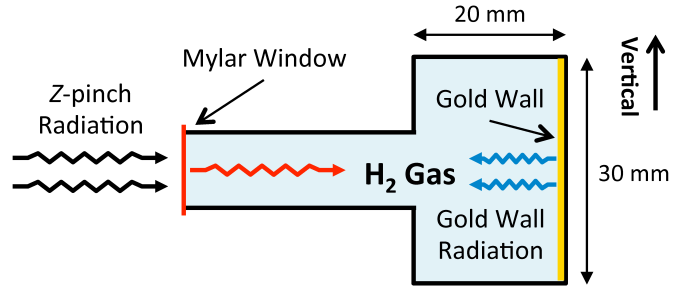}
\caption{Observing line of sight view (perpendicular to incident radiation) schematic of the gas cell cavity cross-section.  {\it Z}-pinch x-rays transmit through the Mylar window and are absorbed by the gold wall at the back end of the cell cavity.  The gold wall re-emits to heat the hydrogen gas.
\label{cell}}
\end{figure}

A brief overview before elaborating in the following subsections: We place a gas cell assembly a distance away from an x-ray source along a radial line of sight (LOS) (Figure \ref{overview}).  X-rays irradiate the gas cell, transmit through a thin ($1.44\pm0.02\,\mu$m) Mylar window, and are absorbed by a gold wall at the back end of the cell cavity (Figure \ref{cell}).  The heated gold re-emits as a Planckian surface with a temperature of $\sim5$\,eV, heating the hydrogen through photoionization.

We observe the hydrogen plasma along lines of sight perpendicular to the {\it z}-pinch radiation and parallel to the gold wall using lens-coupled optical fibers, which deliver the light to time-resolved spectrometer systems.

\subsection{Plasma Formation}\label{formation}

\begin{figure}[]
\includegraphics[width=\columnwidth]{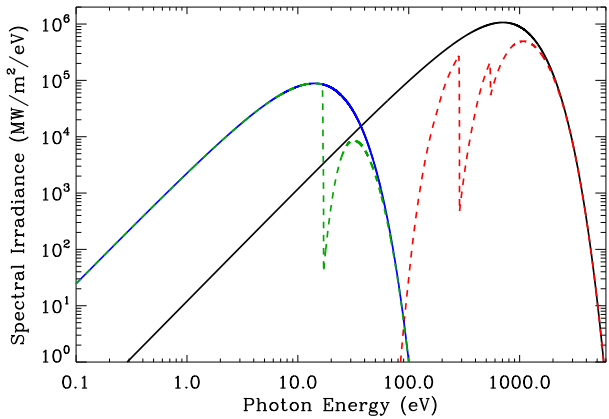}
\caption{Qualitative representation of the radiation environment of the gas cell.  The black curve is the spectral irradiance of the x-ray photons incident on the gas cell.  These photons transmit through the Mylar window as they enter the cell (dashed, red curve).  They are absorbed by a gold wall, which re-emits as a $\sim5$\,eV Planckian (blue curve).  The green curve illustrates how a 10\,mm length of H$_2$ gas at 30\,Torr and at room temperature absorbs the gold wall radiation.\label{rad_env}}
\end{figure}

The source of x-rays incident on the gas cell is a {\it z}-pinch dynamic hohlraum \cite{Spielman98,Sanford02,Rochau08}.  Figure \ref{rad_env} plots the spectral irradiance of the gas cell and within the gas cell cavity.  At peak power, the {\it z} pinch radiates as a $\sim250$\,eV Planckian surface (G. Loisel, private communication).  This Planckian shape is an approximation based on past experiments \cite{Foord04}, and we assume azimuthal uniformity so that each radial LOS observes the same radiation.  The pinch lasts $\sim4$\,ns (full-width at half-maximum of measured absolute power), and its power peaks at $\sim130$\,TW, which yields $\sim1$\,MJ of x-ray energy.  The {\it z}-pinch photons are geometrically diluted as they travel the distance to the gas cell (black curve).  The $\sim1.4\,\mu$m-thick Mylar window attenuates the radiation \cite{Henke93} as it enters the cell (dashed, red curve), eliminating all photons below $\sim80$\,eV.  Hydrogen is transparent to photons at these higher energies; they stream through the gas and do not directly contribute to the plasma heating.  The x-rays are instead absorbed by the gold wall, which then, according to 2D radiation hydrodynamics simulations using LASNEX \cite{Zimmerman78}, re-emits roughly as a $\sim5$\,eV Planckian (blue curve).  Note that the precise spectral distribution is not critical, as long as the gold is emitting photons able to ionize hydrogen.  The blue curve in Figure \ref{rad_env} peaks just above the photoionization edge at 13.6\,eV, where hydrogen becomes opaque.  This helps illustrate that the plasma heating is dominated by the ionizing photons from the gold wall.  Furthermore, at the densities of our experiment we estimate the electron-ion equilibration timescale to be $< 1$\,ns, comparable to our temporal resolution and much shorter than the duration of the experiment.

The amount of energy absorbed by the hydrogen gas is a function of the photon mean free path, which changes as the gas is ionized.  The ionization fraction decreases with increasing distance from the gold wall.  Our preliminary measurements show broader absorption lines (higher ionization) along lines of sight that are closer to the gold, supporting this qualitative picture.

The dashed, green curve in Figure \ref{rad_env} gives a simplistic illustration of the range of photon energies absorbed by the hydrogen.  This particular curve represents transmission through 10\,mm of H$_2$ gas at 30\,Torr and at room temperature \cite{Henke93}.

The plasma formation in the gas cell is similar to that in the boundary region of a Str\"{o}mgren sphere \cite{Stromgren39} surrounding a hot O star.  Both plasmas are predominantly hydrogen, and both sources of ionizing radiation are approximated as blackbodies with $\sim$few\,eV temperature, though the density in our gas cell is significantly higher than even relatively dense regions of the interstellar medium.
 If we calculate an order-of-magnitude Str\"{o}mgren ``distance'' from the gold wall, we find that this balance of photoionization and recombination should occur at $\sim10$\,mm, which is within the dimensions of our gas cell cavity.  The plasma within a Str\"{o}mgren sphere is also isothermal, and, for a hot O star, an electron temperature $T_{\rm e}\sim1$\,eV is typical \cite{Shu91book}.  LASNEX simulations of the environment in our gas cell also predict $T_{\rm e}\sim1$\,eV given a gas fill pressure of 15\,Torr.

\subsection{Gas Cell}

The pulsed power shot configuration for the ZAPP experiments requires a blast shield, which is used to reduce the dissemination of debris resulting from the {\it z} pinch within the the vacuum chamber.  Our gas cell assembly sits immediately outside this blast shield at a distance of $324\pm2$\,mm from the {\it z}-pinch axis.

\begin{figure}[]
\includegraphics[width=\columnwidth]{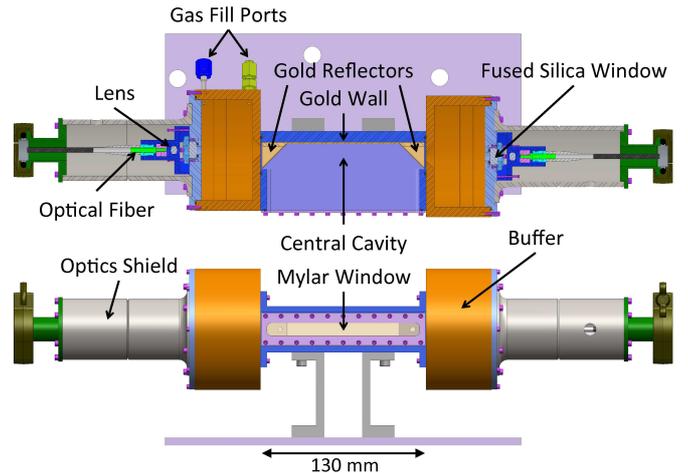}
\caption{Top-view (top) and front-view (bottom) drawings of a gas cell design that uses two lines of sight at different distances from the gold wall and in antiparallel, horizontal directions.  The top-view is in cross-section, revealing gold reflectors used as back-lighting surfaces located on each end of the central cavity.  The buffers (orange) separate the optics from the hot plasma created in this cavity.
\label{drawing}}
\end{figure}

Our gas cell design evolves to enhance the precision of the line shape measurements.  This includes minor modifications to improve robustness, such as tilting the fused silica windows at the ends of the buffers to reduce scattered light from reflections.  It also includes changes to add functionality, such as increasing the number of lines of sight and observing at multiple distances from the gold wall.  Because of this we describe only the major features of the gas cell design that remain consistent through revision.

Figure \ref{drawing} shows a drawing of a design that makes use of two lines of sight.  In this particular design, each LOS observes the plasma at a different distance from the gold wall, which is on the back, inside surface of the cell.  All components are constructed of either 304L stainless steel or 6061-T6 aluminum; gold surfaces are coated onto stainless steel.  The plasma resides in a rectangular cavity that is 120x30x20\,mm.  This cavity is offset 30\,mm beyond the Mylar window (in the radial direction going away from the {\it z} pinch, see Figure \ref{cell}) to extend the lifetime of the experiment by delaying the arrival time of a shock that propagates from the ionized Mylar into our observed LOS.  The central cavity is also flanked by buffers (orange) which provide spatial separation between the optics and the hot plasma.

Following photons from the plasma as they travel to the optical fiber for a given line of sight, they first exit the central cavity through an aperture.  Depending on the particular experiment, this aperture can be in either a thin ($\sim0.3$\,mm) plate whose normal is oriented parallel to the LOS, or in a reflector (gold triangle in top-view) that is angled 45$^\circ$ in the vertical and/or radial directions.  Orienting the gold reflector in this way ensures that this surface has a LOS to the x-rays, which heat it so it can be used as a bright back-lighter, and to the observing optics.  The aperture is typically $4-5$\,mm in diameter, but a different size can be used.  Its function is to limit the ionizing radiation coming from the central cavity and keep the hydrogen gas in the buffer (orange) relatively cool.  The plasma photons travel through this buffer region and through a fused silica window (Newport model 05QW40-30) which interfaces between the hydrogen and the vacuum outside the cell.  With no buffer region, this silica window would be in contact with the hot plasma, thus heating to form its own plasma on its surface and affecting the measurement along the LOS.  Upon exiting the window, a BK~7 bi-convex lens (Newport model KBX010AR.14) or an achromatic doublet lens (Thorlabs model AC060-010-A-ML) focuses the photons onto an optical fiber.  A steel shield (grey) and steel flexible hose protect the optical fiber from high-energy radiation and debris inside the vacuum chamber.

We set the distance between the lens and the fiber so that a beam emerging from the fiber is nearly collimated.  The collection beam diameter is typically $\sim3-4$\,mm but can increase by using a longer focal length lens or decrease by using a limiting aperture.

\subsection{Gas Fill}\label{gasfill}

The gas cell is evacuated ($\lesssim5\times10^{-5}$\,Torr) with the chamber.  We then fill the cell with research purity (99.9999\,\%) H$_{\rm 2}$ gas, making use of a fill and purge procedure repeated three times to flush the system of any lingering contaminants.  Fill pressures range from $3-30$\,Torr, and we measure this pressure {\it in situ} using a piezoresistive pressure sensor (Omegadyne, Inc. PX72-1.5GV).

Assuming the gas fill is at room temperature, the Ideal Gas Law gives us the total particle density with a typical precision of $<2$\,\% (for a fill pressure of $\sim30$\,Torr).  We assume no material escapes the system on the time scale of our experiment ($\frac{1}{2}$\,$\mu$s) so that the total atom density of the gas fill equals the sum of the neutral and ion densities of the created plasma.

\section{Data}

\subsection{Acquisition}

We record time-resolved optical spectra from the lines of sight through our gas cell using two spectrometer systems; more may be added.  Each LOS has its own independent and identical system that collects visible light from the experiment using an optical fiber, transmitting the signal through a spectrometer and to a streak camera with a micro-channel plate intensifier.  The setup is similar to that used previously at {\it Z} \cite{Bailey00,Dunham04,Bailey08}.

A series of four fibers (three connections) link the LOS to the spectrometer.  The first connects the gas cell to a vacuum feed-through port that exits the center chamber housing the ZAPP experiments.  The feed-through fiber connects to a $\sim50$\,m transit fiber bundle which finally connects to a jumper fiber positioned at the input of the spectrometer.  All optical fibers are high OH silica core/clad step-index multimode fibers with a 100\,$\mu$m core diameter and a numerical aperture of $0.22\pm0.02$.

The spectrometer is a 1\,m focal length, $f$/7 aperture Czerny-Turner design (McPherson, Inc. model 2061).  We use 150 or 300\,groove/mm gratings and 70 or 100\,$\mu$m entrance slits.

The streak camera (EG\&G model L-CA-24) sweeps the spectrum over $\sim450$\,ns with $\sim1-2$\,ns temporal resolution.  A micro-channel plate intensifier \cite{LadislasWiza79} amplifies the phosphor emittance exiting the camera, and Kodak T-MAX 400 film records the output.

\subsection{Reduction}

A Perkin-Elmer microdensitometer digitizes the T-MAX 400 film data.  We convert film density to units of radiant exposure (erg cm$^{-2}$) using a NIST-calibrated step wedge.

We set the scale of the time axis using a stream of equally-spaced laser pulses (comb) exposed onto the film during the experiment.  One impulse/comb generator (NSTec Time Mark/Impulse Generator 13) provides these pulses while a second unit provides a single laser pulse (impulse) which we use as a timing fiducial to relate the data to the history of the {\it z} pinch and to data recorded with other spectrometer systems.  The comb also allows us to correct for any time warpage that may occur throughout the streak sweep.

We determine the wavelength dispersion using lasers exposed onto the film prior to the experiment.  These include a 4579\,\AA\ (blue) argon laser (Modu-Laser Stellar Pro 457/4.5) and a 5435\,\AA\ (green) helium neon laser (CVI Melles Griot model 05-LGR-193).  Since the lasers are highly monochromatic ($\Delta\lambda\sim 0.02$--0.03\,\AA), we also recover the instrumental spectral resolution by measuring the width of the lasers in the streak data.  This is typically $\sim9-11$\,\AA\ when using the 150\,groove/mm grating.

\subsection{Example Data}

\begin{figure}[]
\includegraphics[width=\columnwidth]{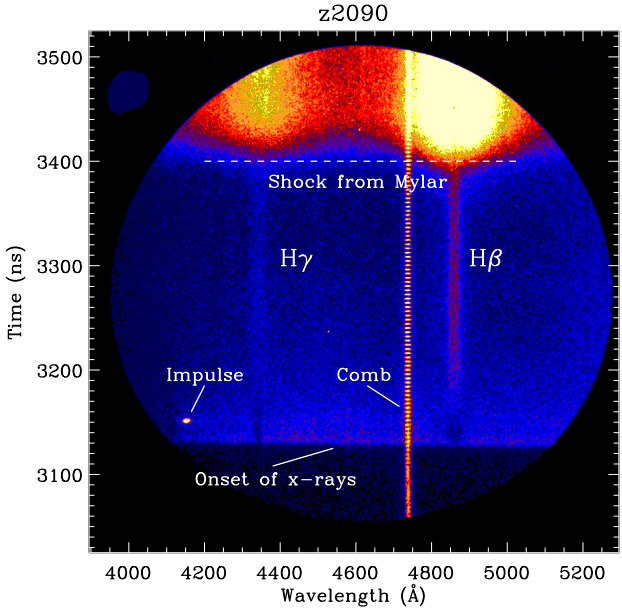}
\caption{Time-resolved spectrum of hydrogen Balmer emission lines from experiment z2090.  After the onset of the x-rays into the gas cell, the plasma forms and reaches a quasi-steady-state in $\sim80$\,ns, which it sustains for $\sim150$\,ns until the shock from the Mylar window reaches our LOS.
\label{emission}}
\end{figure}

\begin{figure}[]
\includegraphics[width=\columnwidth]{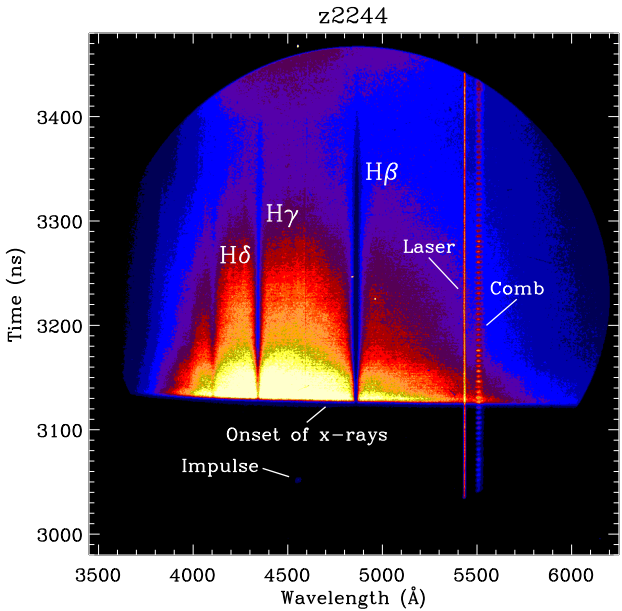}
\caption{Time-resolved spectrum of hydrogen Balmer absorption lines from experiment z2244.  We can observe higher $n$ lines in absorption due to the increased signal-to-noise.  The intensity of the back-lighting continuum decreases over the lifetime of the experiment as the gold reflector cools.
\label{absorption}}
\end{figure}

Figures \ref{emission} and \ref{absorption} show time-resolved spectra (streak images) of hydrogen plasmas from two different experiments.  In both streak images, the hydrogen gas is invisible until the {\it z}-pinch x-rays arrive to heat the gold wall.  Within $\sim80$\,ns from the onset of the x-rays, the plasma reaches a quasi-steady-state.  This is more easily observed in the emission data (Figure \ref{emission}) where the Balmer lines sustain a constant width and intensity for $\sim150$\,ns until the shock from the Mylar window (heated by the x-rays) reaches our LOS.  The back-lighting continuum in the absorption data (Figure \ref{absorption}) decreases in time as the gold reflector cools; this changes the contrast between the continuum and the Balmer lines, making it difficult to notice by eye that the line width in absorption remain constant through time.  It is important to note that the signal-to-noise significantly increases when we observe the lines in absorption.  This allows us to measure higher principal quantum number ($n$) lines than in the emission case.

These data are not corrected to flatten the instrumental sensitivity or to account for the wavelength-dependent transit time and attenuation through the optical fibers \cite{Cochrane01}.

\section{Discussion}

\subsection{Experimental Context}

What distinguishes this experiment from others in the past that have investigated hydrogen Balmer line profiles in the range of conditions 
$T_{\rm e}\sim 1$\,eV and $n_{\rm e}\sim 10^{17}$\,cm$^{-3}$ is the plasma source.  Here it is radiation-driven.  This decouples the heating source from the plasma we are studying in the sense that the initial conditions of the hydrogen gas, such as the fill pressure, have no effect on the radiation temperature of the gold wall.  The gold temperature can be increased, for example, by moving the gas cell closer to the {\it z} pinch or by using a thinner Mylar window.  This ``decoupling'' give us potentially finer control of the experimental conditions as well as the ability to explore a broad range of plasma conditions.

\subsubsection{Achievable Range of Electron Densities}

\begin{figure}[]
\includegraphics[width=\columnwidth]{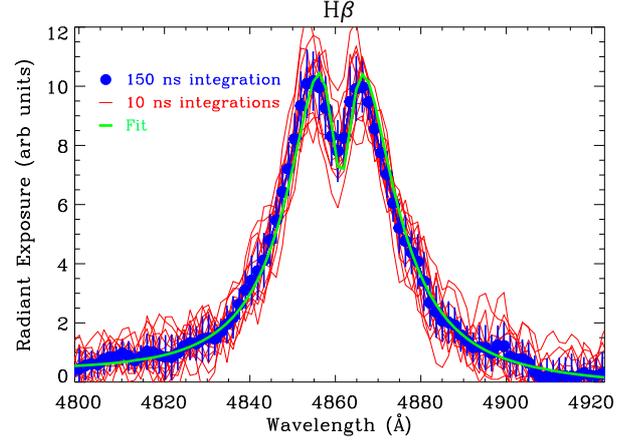}
\caption{H$\beta$ emission line profiles integrated over 10\,ns durations (red curves) throughout the quasi-steady-state period (3215--3365\,ns) from the streak image in Figure \ref{emission}.  The combined 150\,ns integration curve (filled, blue circles) includes the standard deviation of the shorter time integrations, illustrating the level to which they the stay similar.  We estimate $n_{\rm e}\sim6\time 10^{16}$\,cm$^{-3}$ by fitting this measured profile to the tabulated profiles from \citet{Lemke97} (green curve) which uses the theory of \citet{Vidal73}.
\label{shape_in_time}}
\end{figure}

We estimate electron density $n_{\rm e}$ by fitting our observed H$\beta$ spectral lines to the tabulated profiles from \citet{Lemke97}; these use the line broadening theory of \citet{Vidal73}.  Our preliminary fits do not yet account for optical depth effects, which can be significant in the line center and become negligible farther out in the line wings.  Figure \ref{shape_in_time} plots an example fit (green curve) to a 150\,ns integration (filled, blue circles) of the H$\beta$ emission line from experiment z2090 (Figure \ref{emission}).  Here $n_{\rm e}\sim 6\times 10^{16}$\,cm$^{-3}$.  Figure \ref{shape_in_time} also plots the constituent 10\,ns integrations (red curves) of the full 150\,ns-integrated H$\beta$ line.  The standard deviation of these red curves divided by the square root of the number of curves is a measure of the experimental uncertainty for the spectrum.  The signal-to-noise ratio (S/N), estimated this way, at half of the maximum exposure of this H$\beta$ emission line is $\gtrsim28$ for a single spectral element.

\begin{figure}[]
\includegraphics[width=\columnwidth]{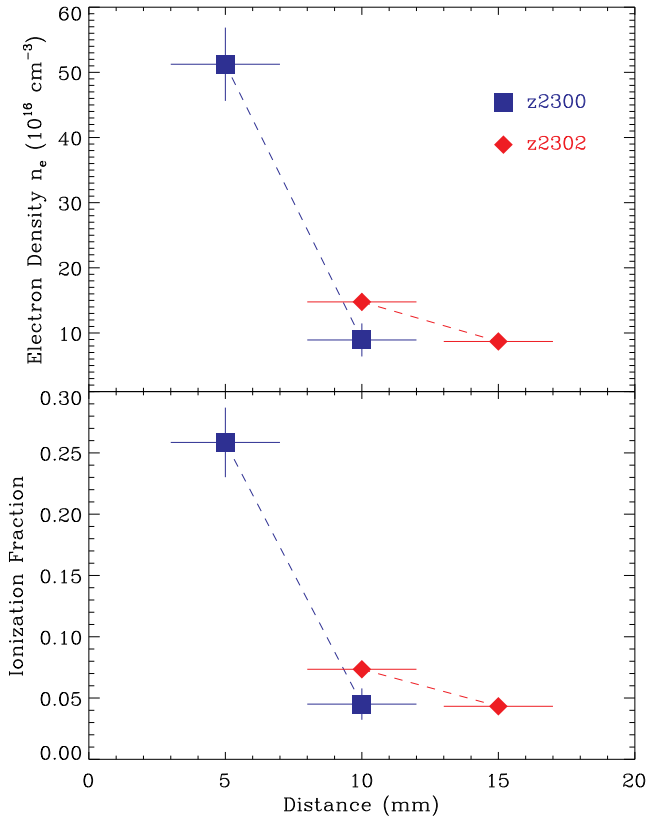}
\caption{Estimates of electron density $n_{\rm e}$ (top panel) and ionization fraction (bottom panel) of our hydrogen plasma as a function of distance from the gold wall.  Data points are from H$\beta$ absorption line fits to spectra integrated over 40\,ns.  The vertical lines are the standard deviations of fit values to 2\,ns integrations making up the full integrations, and the horizontal lines are the nominal diameters of the observed lines of sight.  Dashed lines connect data points sampling lines of sight from the same experiment.
\label{distance}}
\end{figure}

Close to the gold wall, the hydrogen is nearly fully ionized.  Farther from the gold, ionization decreases.  We demonstrate this in Figure \ref{distance}, which shows data from two different experiments that use the gas cell design illustrated in Figure \ref{drawing}.  In experiment z2300, we simultaneously observe the H plasma in absorption along a LOS that is 5\,mm away from the gold wall and a LOS that is 10\,mm away.  For experiment z2302, the lines of sight are at distances of 10\,mm and 15\,mm from the gold wall.  Each data point in the top panel of Figure \ref{distance} is the fit value of $n_{\rm e}$ for a 40\,ns integration of the H$\beta$ absorption line from streak data.  These integrated spectra sample the same time during the life of the plasma for lines of sight in the same experiment (connected by dashed lines).  Between the two experiments, they sample the same time with respect to plasma formation, starting $\sim24$\,ns after the onset of x-rays.  The solid, vertical lines quantify the stability of $n_{\rm e}$; these are the standard deviation of fit values of 20 integrations, each 2\,ns long, occurring throughout the full 40\,ns integration.  The solid, horizontal lines are the nominal diameters (4\,mm) of the observed lines of sight, and the entire scale of the horizontal axis is equal to the size of the gas cell cavity.

We also show the ionization fraction ($\frac{n_{\rm e}}{n_{\rm H}+n_{\rm H^+}}$; recall that our plasma is hydrogen so that $n_{\rm e}=n_{\rm H^+}$) as a function of distance from the gold wall (bottom panel).  The gas fill pressures for z2300 and z2302 are $30.42\pm0.72$\,Torr and $30.79\pm0.69$\,Torr, respectively, which correspond to $n_{\rm H}+n_{\rm H^+} = (1.982\pm0.024)\times 10^{18}$\,cm$^{-3}$ and $(2.010\pm0.023)\times 10^{18}$\,cm$^{-3}$.  Note that the electron density (ionization fraction) falls off less steeply farther from the gold wall.

The upper limit of the range of $n_{\rm e}$ we can achieve is set by the gas fill pressure (in the fully ionized scenario).  For a fill pressure of 30\,Torr of H$_2$, this is $n_{\rm e}\sim 2\times 10^{18}$\,cm$^{-3}$.  We can comfortably increase the fill pressure by roughly a factor of two.  Beyond that is feasible, but we would need to take better care in securing the Mylar window against leaks and rupture.

In principle we can achieve arbitrarily low values of $n_{\rm e}$ (as the ionization fraction goes to zero).  What is measureable and useful for line profile studies after considering line strength and S/N, however, does have a lower limit.

The highest $n_{\rm e}$ we encounter and can demonstrate to be useful thus far is $n_{\rm e}\sim6\times 10^{17}$\,cm$^{-3}$.  The lowest is in the range of $n_{\rm e}\sim2\times 10^{16}$\,cm$^{-3}$.  These come from fits to H$\beta$ absorption lines.  For a typical total neutral and ion density $n_{\rm H}+n_{\rm H^+}\sim2\times 10^{18}$\,cm$^{-3}$, these electron densities correspond to ionization fractions of $\frac{n_{\rm e}}{n_{\rm H}+n_{\rm H^+}}\sim0.27$ and $\sim0.01$, respectively.

\subsubsection{Comparison With Other Experimental Platforms}

Forming a hydrogen plasma using radiation requires a sufficient number of ionizing photons distributed over a relatively large surface area.  This is only possible with a large input of energy to the experimental system, which, in our case, is supplied by the large amount of x-rays from a {\it z}-pinch dynamic hohlraum.  

To understand how our radiation-driven plasma source compares with other plasma sources, we look to the discussion of \citet{Konjevic76a,Konjevic76b,Konjevic99} and \citet{Konjevic02}.  Though these reviews do not include experiments investigating hydrogen specifically, they provide an excellent contextual description of the plasma sources, diagnostic methods, and experimental considerations pertinent for laboratory studies of Stark-broadened spectral lines.

As pointed out in \citet{Wiese72}, the two most desirable properties of a plasma source are that it is stationary (in a steady-state of plasma conditions) and homogeneous.  None are truly stationary or homogenous, but different sources can achieve either or both of these properties to some degree.

Shock-heated experiments 
\cite[e.g.,][]{Doherty55,Berg62,McLean65,Bengtson69,Hey75,Okasaka77,Djurovic05} observe a plasma in a shock front that travels through a tube.  If the shock front is planar and the boundary layers negligible, then one achieves a homogeneous plasma by observing along a LOS that is tangential to the plane of the shock front.  Since this plasma is not stationary, the temporal resolution of the observation must be sufficient.  The experimental setup must also be highly reproducible to repeat the experiment many times to accumulate enough signal.  In the published literature, hydrogen has not always been the primary component of the gas fill.

One may also achieve homogeneity using pulsed discharges 
\cite[e.g.,][]{Vujnovic62,Hill67,Morris68,Torres84} by appropriately arranging the observing optics (i.e., by observing a discharge plasma ``end-on'' along its axis) and while paying special attention to the boundary layers.  These plasmas are also not stationary and must be highly reproducible.  Again, hydrogen has not always been the primary component of the gas fill.

Stabilized arc experiments are stationary, which is advantageous for data collection and for achieving high signal-to-noise data.  These plasmas are not radially homogeneous, so when they are observed ``side-on'' (perpendicular to the arc axis) \cite[e.g.,][]{Wiese63,Shumaker68,Wiese72,Ershov-Pavlov87}, one must correct for the radial temperature distribution using an Abel inversion.  Some investigations observe the plasma ``end-on'' 
\cite[e.g.,][]{Baessler80,Helbig81,Halenka86,Djurovic88}, but, like the pulsed discharge experiments, properly addressing the boundary layers in the plasma is a significant difficulty.  Also, with the exception of \citet{Wiese63} and \citet{Wiese72}, the experiments we list here operate arcs in gases that are mostly argon with only small amounts of hydrogen.  The reason is for plasma diagnostics using Ar spectral lines as well as to observe the hydrogen at desired optical depths.  However, the shapes of the hydrogen profiles are affected by Ar lines due to blending.  Removing these lines introduces a systematic uncertainty.

Our discussion does not include all possible plasma sources.  We attempt to cover the ones most relevant to our investigation.  For example, though experiments using radio frequency discharge plasmas \cite[e.g.,][]{Schluter66,Bengtson70} observe many high $n$ lines, this is because they are low density, and this is too low for our interests.  Laser-driven optical discharge plasmas \cite[e.g.,][]{Carlhoff86,Uhlenbusch90} are stationary, but they are too high density, leaving only the lowest $n$ Balmer lines separated enough to be useful.  Laser-induced optical breakdown plasmas span a large range of conditions but are transient, inhomogeneous, and spatially small \cite{Parigger95b,Parigger03}.

The plasma generated in our experiment is not as stationary as the stabilized arc plasmas, but as we mention earlier and point out in Figure \ref{emission}, it does reach a quasi-steady-state where the Balmer line shapes remain constant for some time.  We illustrate this in Figure \ref{shape_in_time}, which plots 10\,ns integrations of the H$\beta$ emission line (red curves) throughout the quasi-steady-state period (3215--3365\,ns) from Figure \ref{emission}.  The standard deviation of these red curves, which we use to estimate S/N, is also a measure of the stability of the spectral line throughout time.

Our plasma is also heated by radiation coming off of a relatively large (120x10\,mm) planar surface.  This reduces the limitations of observation imposed by the localized heating that occurs with pulsed discharges or arcs.  We are not concerned with accurately integrating our observed LOS over an arc or annulus of constant radius about an axis as is needed for ``end-on'' observations.  Neither are we concerned with performing Abel inversions for ``side-on'' observations.  We are free to change the distance of our LOS from the gold wall, thus probing different ionizations within the same plasma.  The large heating surface allows us to observe longer path lengths than can be achieved in other plasma sources, and we can vary these lengths more readily.  Observing longer path lengths means we can measure spectral lines that are more optically thin without having to increase the density or integrate over long exposure times.  The large surface also aides us in achieving homogeneity (along lines of sight at a constant distance from the wall) by minimizing the sensitivity of the plasma heating to small inhomogeneities in the gold wall temperature.

Perhaps the most distinguishing feature of our experiment is the ability to observe hydrogen Balmer lines in absorption in addition to emission.  Just as x-rays from the {\it z}-pinch heat the gold wall at the back end of the gas cell cavity, they also heat another gold surface that is angled so as to have a clear line of sight to both the observing optics and the {\it z}-pinch radiation.  The re-emission from the gold provides a bright continuum background useful for absorption measurements across visible wavelengths.

It is interesting to note that, because experiments have never before had such a back-lighter available that allowed for absorption measurements, there has never been a convention in line profile studies to specify emission versus absorption line profiles.  ``Line profile'' has always implied ``{\it emission} line profile''.  In fact, one must search the literature carefully in order to encounter this more precise jargon, but even then, it is used in a hybrid form in the context of self-absorption of an emission line \cite[e.g.,][]{Burgess72}.   We also point out that the spectral lines in WD observations, from which atmospheric parameters are determined, are all measured in absorption.

\subsection{Strategy for Addressing the White Dwarf Problem}

\begin{figure}[]
\includegraphics[width=\columnwidth]{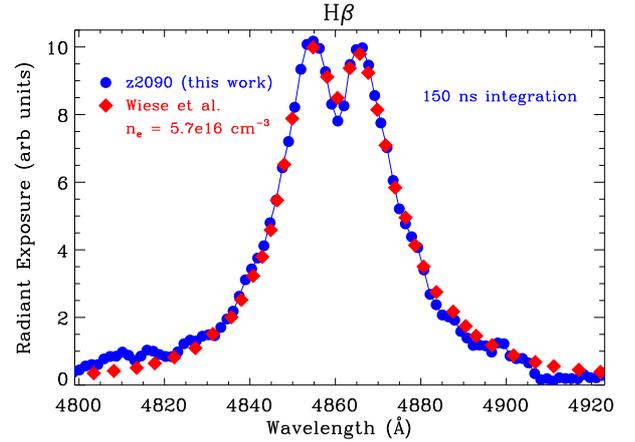}
\caption{H$\beta$ spectrum (filled, blue circles) integrated over the 150\,ns quasi-steady-state period (3215--3365\,ns) from the streak image in Figure 
\ref{emission}.  The filled, red diamonds are the data from the stabilized arc experiment of \citet{Wiese72}, which determines an electron density 
$n_{\rm e} = 5.7\times10^{16}$\,cm$^{-3}$.
\label{wiese}}
\end{figure}

Independent of $n_{\rm e}$ estimates from line fitting, we demonstrate that we are achieving our target plasma conditions in Figure \ref{wiese}, which plots an H$\beta$ emission line profile from experiment z2090 with the same line from the wall-stabilized arc experiment by \citet{Wiese72}.  We regard the work by \citet{Wiese72} as the standard for hydrogen Balmer line profiles at these conditions because of the high quality of their data and because of the influence it has on theoretical work in astrophysics 
\cite[e.g.,][]{Hummer88,Rohrmann02,Tremblay09}.  An initial step in the application of our experimental results to the white dwarf problem described in Section \ref{intro} will be a comparison amongst line shapes from \citet{Wiese72} and other relevant experiments.

Ideally we want to measure the conditions of our plasma using methods that are independent from spectral line shapes.  This may include laser interferometry or optical Thomson scattering techniques.  We currently do not have these capabilities, but while they are in development, we can still make significant progress in addressing our scientific goals.

We will measure relative line shapes for H$\beta$, H$\gamma$, and H$\delta$ (and H$\epsilon$ if we achieve sufficient S/N) formed in a single, pure hydrogen plasma at WD photospheric conditions.  Since the H$\beta$ line shape is in good agreement between theories, we can use it as a plasma diagnostic and as a basis to compare higher $n$ Balmer lines.  The discrepancy between theories increases with higher principal quantum number, and it is the higher $n$ lines that are more sensitive to the surface gravity (log\,$g$) in WD atmospheres.  The ultimate implementation of our experiment will use additional diagnostic capabilities to further constrain our plasma.

\subsection{Scientific Potential}

Our experimental platform may also explore other gas compositions, forming these plasmas similarly to how hydrogen is formed \cite{Falcon12c}.  The next most relevant species for WD atmospheres is helium \cite[e.g.,][]{Voss07,Bergeron11}.  Like the hydrogen-atmosphere WDs, helium-atmosphere WDs also suffer from an apparent increase in log\,$g$ with decreasing effective temperature \cite{Kepler07,Bergeron11}.

There are also the carbon-atmosphere WDs, discovered only a few years ago \cite{Dufour07}.  Initial spectroscopic fits to the observed spectra were poor \cite{Dufour08}, prompting new calculations of Stark-broadened C~II line profiles \cite{Dufour11}.  Fits incorporating these newer theoretical profiles suggest significantly different atmospheric parameters for the same stars, and further calculations for O~I, O~II, and C~I are underway \cite{Dufour11}.  The timing is ripe for laboratory support for all these calculations.

This platform is not limited to the study of WD atmospheres.  Our experimental plasma is photoionized, and photoionized plasmas exist in many forms throughout astrophysics.  Due to the inherent difficulty of producing a sufficient number of ionizing photons to create significant volumes of plasma, however, these kinds of laboratory experiments are sparse \cite{Mancini09}.  One obvious area to expand this research is into the investigation of non-LTE, time-dependent collisional-radiative atomic kinetics.

\section{Conclusions}

We describe an experimental platform to create hydrogen plasmas in the range of conditions that exist in white dwarf photospheres 
($T_{\rm e}\sim1$\,eV, $n_{\rm e}\sim10^{17}$\,cm$^{-3}$).  Here we measure hydrogen Balmer line profiles in emission and, for the first time, in absorption.  These profiles will be used to constrain the latest theoretical WD atmosphere models, which, when used with the spectroscopic method, are responsible for determining fundamental parameters (i.e., effective temperature, mass) for tens of thousands of WDs.  We perform our experiment at the {\it Z} Pulsed Power Facility at Sandia National Laboratories, making use of its powerful x-ray capability to drive plasma formation in a gas cell.  Our plasma source is radiation-driven, which is unique compared to past experiments and which decouples the heating source from the plasma to be studied.

\section*{Acknowledgments}
This work was performed at Sandia National Laboratories and is supported by the Laboratory Directed Research and Development program.  We thank the {\it Z} dynamic hohlraum, accelerator, diagnostics, materials processing, target fabrication, and wire array teams, without which we cannot run our experiments.  In particular we thank C. Ball, R. Bengtson, I. Hall, D. Headley, M. Jones, N. Joseph, P. Lake, T. Lockard, G. Loisel, R. Mancini, Y. Maron, T. Nagayama, L. Nielsen-Weber, D. Sandoval, K. Shelton, T. Strizic, M. Vigil, and J. Villalva.  Sandia is a multiprogram laboratory operated by Sandia Corporation, a Lockheed Martin Company, for the United States Department of Energy under contract DE-AC04-94AL85000.  This work has made use of NASA's Astrophysics Data System Bibliographic Services.  R.E.F. acknowledges support of the National Physical Science Consortium, and R.E.F., M.H.M., and D.E.W. gratefully acknowledge support of the Norman Hackerman Advanced Research Program under grant 003658-0252-2009.



\bibliographystyle{elsarticle-num-names}
\bibliography{/home/grad79/cylver/all}


\end{document}